\documentclass[12pt, notitlepage]{article}
\pdfoutput=1

\usepackage[toc,page]{appendix}

\usepackage{float}
\usepackage{subfig}
\usepackage{graphicx}
\usepackage[T1]{fontenc}
\usepackage[utf8]{inputenc}
\usepackage{authblk}
\usepackage{amsmath,bm}
\usepackage[left=3cm,right=3cm,top=2cm,bottom=2cm]{geometry}
\renewcommand\footnotemark{}

\usepackage{color}

\begin{document}

\title{New higher-order transition in causal dynamical triangulations}         

\author[a,b]{J.~Ambjorn}
\author[a,c]{D.~Coumbe}
\author[c]{J.~Gizbert-Studnicki}
\author[c]{A.~G\"orlich}
\author[c]{J.~Jurkiewicz}
\affil[a]{\small{The Niels Bohr Institute, Copenhagen University, \authorcr Blegdamsvej 17, DK-2100 Copenhagen Ø, Denmark. \authorcr E-mail: ambjorn@nbi.dk, daniel.coumbe@nbi.ku.dk.\vspace{+2ex}}} 

\affil[b]{\small{IMAPP, Radboud University, \authorcr Nijmegen, PO Box 9010, The Netherlands.\vspace{+2ex}}}

\affil[c]{\small{The M. Smoluchowski Institute of Physics, Jagiellonian University, \authorcr \L ojasiewicza 11, Krak\'ow, PL 30-348, Poland. \authorcr Email: jakub.gizbert-studnicki@uj.edu.pl, andrzej.goerlich@uj.edu.pl, jerzy.jurkiewicz@uj.edu.pl.}}


\date{\small({Dated: \today})}          
\maketitle


\begin{abstract}
We reinvestigate the recently discovered  bifurcation phase transition in Causal Dynamical Triangulations (CDT) and provide further evidence that it is  a higher order transition. We also investigate the impact of introducing matter in the form of massless scalar fields to CDT. We discuss the impact of scalar fields on the measured spatial volumes and fluctuation profiles in addition to analysing how the scalar fields influence the position of the bifurcation  transition.

\vspace{1cm}

\end{abstract}


\begin{section}{Introduction}\label{intro}

The reasons for attempting to quantize gravity are manifold, including the fact that every other fundamental force can be understood within the framework of quantum field theory. However, treating gravity as a perturbative quantum field theory results in a complete loss of predictive power, since in order to define such a theory one would first need to experimentally determine an infinite number of independent coefficients. The divergent number of counterterm coefficients associated with the perturbative treatment of general relativity have been confirmed by explicit calculation, appearing at two-loops for pure gravity~\cite{Goroff:1985th} and at one-loop for gravity including matter~\cite{'tHooft:1974bx}. The divergences associated with the perturbative treatment of gravity has generated considerable interest in nonperturbative formulations, one of the most promising of which is the so-called asymptotic safety scenario.

First proposed by Weinberg~\cite{Weinberg79}, asymptotic safety posits the existence of an ultra-violet fixed point (UVFP) under the flow of the renormalization group of gravitational couplings. If there exist only a finite number of such couplings that are attracted to the fixed point at high energies, then asymptotic safety may define a finite and predictive theory of quantum gravity in the nonperturbative regime. There is mounting evidence for the existence of an UVFP, ranging from the $(2+\epsilon)$-expansion of spacetime dimensionality~\cite{Christensen:1978sc} to functional renormalization group results~\cite{Benedetti:2009rx,Codello:2008vh,Litim:2003vp}. A lattice formulation of quantum gravity provides a complimentary approach to asymptotic safety, since it permits the definition of a gravitational path integral that can be studied in the nonperturbative regime. Lattice gravity can also provide direct evidence for asymptotic safety, since in a lattice formulation an UVFP would appear as a higher (than first) order critical point, the approach to which would define a continuum limit. 

One of the first lattice regularizations of quantum gravity is Euclidean dynamical triangulations (EDT), which attempts to define a nonperturbative theory of quantum gravity as the continuum limit of a sum over discrete spacetime geometries. In this approach spacetime is approximated by a network of locally flat $d$-dimensional triangles that are connected via their $(d-1)$-dimensional faces. Unfortunately, early EDT simulations found just two phases, neither of which resembled $4$-dimensional semiclassical general relativity.\interfootnotelinepenalty=10000 \footnote{\scriptsize Although there are some encouraging signs that a particular modification of EDT may have a suitable infra-red limit after a certain fine-tuning is implemented~\cite{Laiho:2016nlp}.} Moreover, it was shown that these two phases are separated by a first order phase transition, making the existence of a continuum limit improbable. Motivated by the difficulties encountered in the original EDT formulation, a causality condition was added to the model whereby the lattice is foliated into space-like hypersurfaces of fixed topology, an approach known as causal dynamical triangulations (CDT)~\cite{Ambjorn:1998xu}. The inclusion of this additional constraint appears to cure the problems found in the original EDT formulation. 


The path integral for pure CDT quantum gravity is defined by

\begin{equation} \label{eq:CDTPartitionFunction}
Z_{E}={\sum_{T}}\frac{1}{C_{T}}e^{-S_{EH}(T)},
\end{equation}

\noindent where one performs a sum over all discrete triangulations $T$ allowed by the causality constraint. $C(T)$ is a symmetry factor encoding the number of equivalent ways of labelling the vertices in $T$, and $S_{EH}(T)$ is the discretised Einstein-Hilbert action of the triangulation~\cite{Regge:1961px}, where

\begin{equation} \label{eq:GeneralEinstein-ReggeAction}
S_{EH}(T)=-\left(\kappa_{0}+6\Delta\right)N_{0}+\kappa_{4}\left(N_{4,1}+N_{3,2}\right)+\Delta\left(2N_{4,1}+N_{3,2}\right).
\end{equation}

\noindent $N_{i,j}$ denotes the number of simplicial building blocks with $i$ vertices on hypersurface $t$ and $j$ vertices on hypersurface $t+1$. The number of vertices in the triangulation is given by $N_0$. The CDT action includes three bare coupling constants $\kappa_{0}$, $\Delta$ and $\kappa_{4}$. $\kappa_{0}$ is inversely proportional to Newton's constant, $\Delta$ is related to the ratio of the length of space-like and time-like links on the lattice and $\kappa_{4}$ is proportional to the cosmological constant. $\kappa_{4}$ is tuned to a (pseudo)-critical value in the simulations such that one can take an infinite-volume limit. The parameter space of CDT can then be explored by varying $\kappa_{0}$ and $\Delta$.


Using Monte Carlo simulations the CDT parameter space spanned by $\kappa_{0}$ and $\Delta$ has now largely been mapped out, as shown schematically in Fig.~\ref{pdnew}. To date there are 4 known phases of CDT, labelled A, B, $C_{dS}$ and $C_{b}$. Phases $A$ and $B$ do not appear to reproduce general relativity in the semiclassical limit, and are generally regarded as lattice artifacts. The recently discovered bifurcation phase $C_{b}$ also has a number of unphysical features~\cite{Ambjorn:2016mnn} such as a very large, and possibly infinite, effective spacetime dimension~\cite{Ambjorn:2015fza}. However, the de Sitter phase $C_{dS}$ has a volume profile that closely matches Euclidean de Sitter space~\cite{Ambjorn:2008wc} and an effective dimension consistent with $4$~\cite{Ambjorn05,Ambjorn:2005db,Coumbe:2014noa}, thus defining the physically interesting phase of CDT.  

\begin{figure}[H]
\centering
\includegraphics[width=0.6\linewidth]{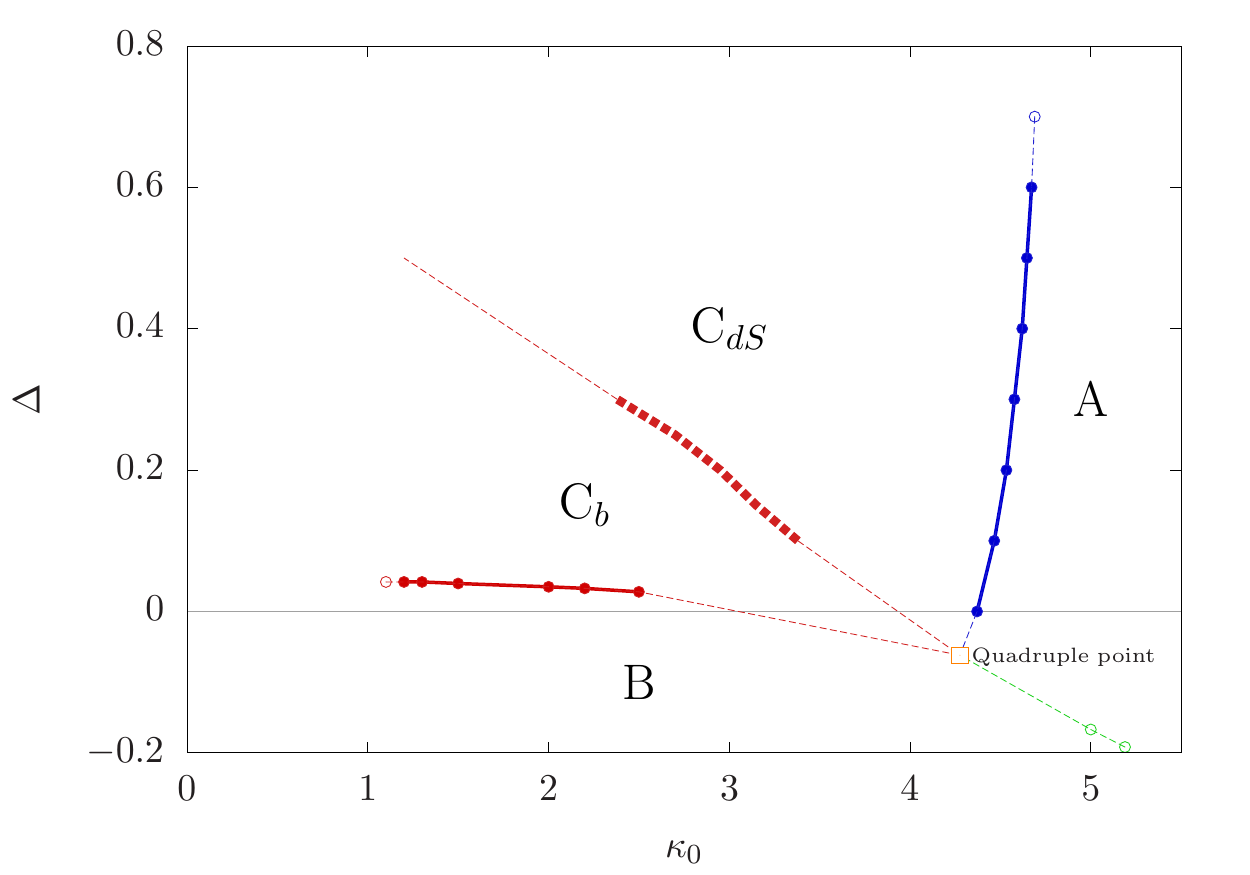}
\caption{\small The updated phase structure of $4$-dimensional CDT. }
\label{pdnew}
\end{figure}

The $A-C_{dS}$ transition is known to be first order, while the $B$-$C_{b}$ transition is likely second order~\cite{Ambjorn:2012ij}. The order of the $C_{dS}-C_{b}$ transition has yet to be definitively determined, although preliminary calculations suggest a higher-order transition~\cite{Coumbe:2015oaa}. Conclusively determining the order of the $C_{dS}-C_{b}$ transition may prove an important result in CDT, since a second order transition would raise the possibility of defining a continuum limit from within the physically interesting de Sitter phase $C_{dS}$. In this work we aim to more definitively determine the order of the $C_{dS}-C_{b}$ transition.  

Another question that this work addresses is how the inclusion of matter fields affects the phase structure of CDT. In particular, it is possible that the bifurcation phase $C_{b}$ is an artifact of the naive pure gravity formulation of CDT, and that the inclusion of a sufficient number of matter fields may be a necessary condition for the universe to exhibit the correct semiclassical behaviour, as suggested in Ref.~\cite{Hartle:2008ng}. In this work we investigate this possibility by coupling CDT to $N$ massless scalar fields and examining how this affects the extent of the bifurcation phase. 

This paper is organised as follows. In section~\ref{OrderOfTrans} we define the order parameter used to study the $C_{dS}-C_{b}$ phase transition and detail how a finite-size scaling analysis can be used to indicate the order of this transition. After reviewing some technical details in section~\ref{fields} regarding the numerical implementation of adding massless scalar fields to CDT, we study how the inclusion of such matter fields affects the position of the $C_{dS}-C_{b}$ transition in the parameter space. A discussion and summary of the results obtained in this work are presented in section~\ref{Discussion}.  

\end{section}

\begin{section}{The order of the $C_{dS}-C_{b}$ transition}\label{OrderOfTrans}

Phase transitions are often associated with a breaking of some symmetry. To quantify the transition one can define an order parameter $OP$ that captures the symmetry difference between the phases. Such an order parameter is typically zero (or constant) inside the symmetric phase and non-zero (or non-constant) in the symmetry broken phase. The first order phase transition point is then characterised by a discontinuity in the first order derivative of the $OP$ in the infinite volume limit, whereas for a $n$th order transition the $1,...,(n-1)$th order derivatives are continuous but the $n$th order derivative is not.  

In numerical simulations it is quite difficult to distinguish the order of a phase transition by just looking at the (dis)continuity of some order parameter's $n$th derivative. The reason is twofold. First, numerical simulations are always  performed with  finite precision and it is very difficult to judge whether a sudden jump in some order parameter (or its derivative) is caused by a real discontinuity, or is actually caused by insufficient measurement precision. Secondly, numerical simulations always require finite systems (volumes) and thus no real phase transitions take place (all infinities are replaced by large but finite numbers dependent on the system's size). One should therefore carefully analyse finite size effects and extrapolate the results to the infinite volume limit. 

As an example, one can  locate (pseudo-)critical points by searching the parameter space for peaks in the susceptibility of the order parameter
\begin{equation}\label{susceptibility}
\chi_{OP} = \langle OP ^2 \rangle - \langle OP \rangle^2.
\end{equation}
Positions of such transition points in the parameter space will typically depend on the system volume and by measuring how they change with increasing volume one can in principle determine  the position of the true phase transition in the infinite volume limit by extrapolation. One can also use the same method to determine critical exponents, whose values may indicate the order of the phase transition.

The results presented in this section are a continuation of work initiated in Refs.~\cite{Ambjorn:2015qja,   Coumbe:2015oaa}, where a suitable choice of order parameters was suggested based on  microscopic geometric properties of the bifurcation phase $C_{b}$.  Distribution of volume in  phase $C_{b}$ is markedly different than in the de Sitter phase $C_{dS}$, with spatial volume concentrated in clusters connected by vertices of very high coordination number \cite{Ambjorn:2015qja, Ambjorn:2016mnn}. This geometric difference is presumably caused by a breaking of  homogeneity\footnote{The homogeneity of phase $C_{dS}$ should be understood in a statistical sense, i.e. the emergent average semiclassical background geometry is homogeneous but individual trajectories of the path integral (triangulations) are not. {This is in analogy with the ordinary path integral of quantum mechanics where the classical (average) trajectory is smooth but individual path integral trajectories are nowhere  differentiable.} This homogeneity is not the case in phase $C_b$ where the average geometry is not homogeneous and also individual triangulations are much less homogeneous than inside phase $C_{dS}$.  {This is due to the formation of large volume clusters around vertices of very high coordination number present every second lattice time coordinate in phase $C_b$ (see Fig. \ref{nocoord}, right, and Fig. \ref{Omax}). Such volume clusters constitute most of the individual triangulations of phase $C_b$ and they overlap forming a four-dimensional structure.  Geometry inside the structure is markedly different than the geometry outside since neither the  average space-time geometry nor average spatial geometries are homogeneous. In phase $C_{dS}$ some volume clusters also form around the highest order vertices due to quantum fluctuations, however in this case the volume clusters are much smaller and their overlap is only statistical. As a result there is no distinct four-dimensional structure and the average space-time and spatial geometries seem to be  homogeneous. The differences between the geometry of phases $C_b$ and  $C_{dS}$ will be discussed in detail in forthcoming articles.   
}
} of phase  $C_{dS}$ and it 
can be exploited to signal the transition  to  phase $C_{b}$.

\begin{figure}[H]
\centering
\includegraphics[width=0.45\linewidth]{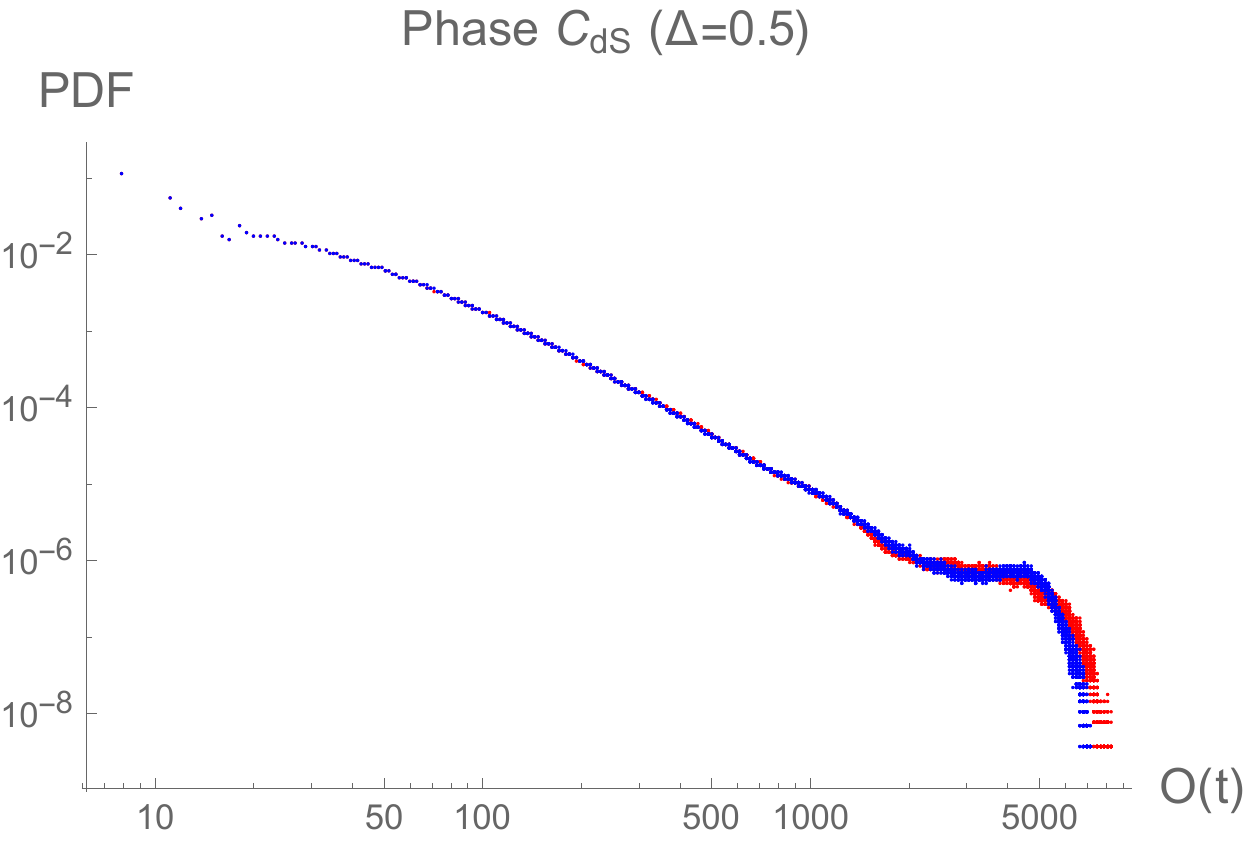}
\includegraphics[width=0.45\linewidth]{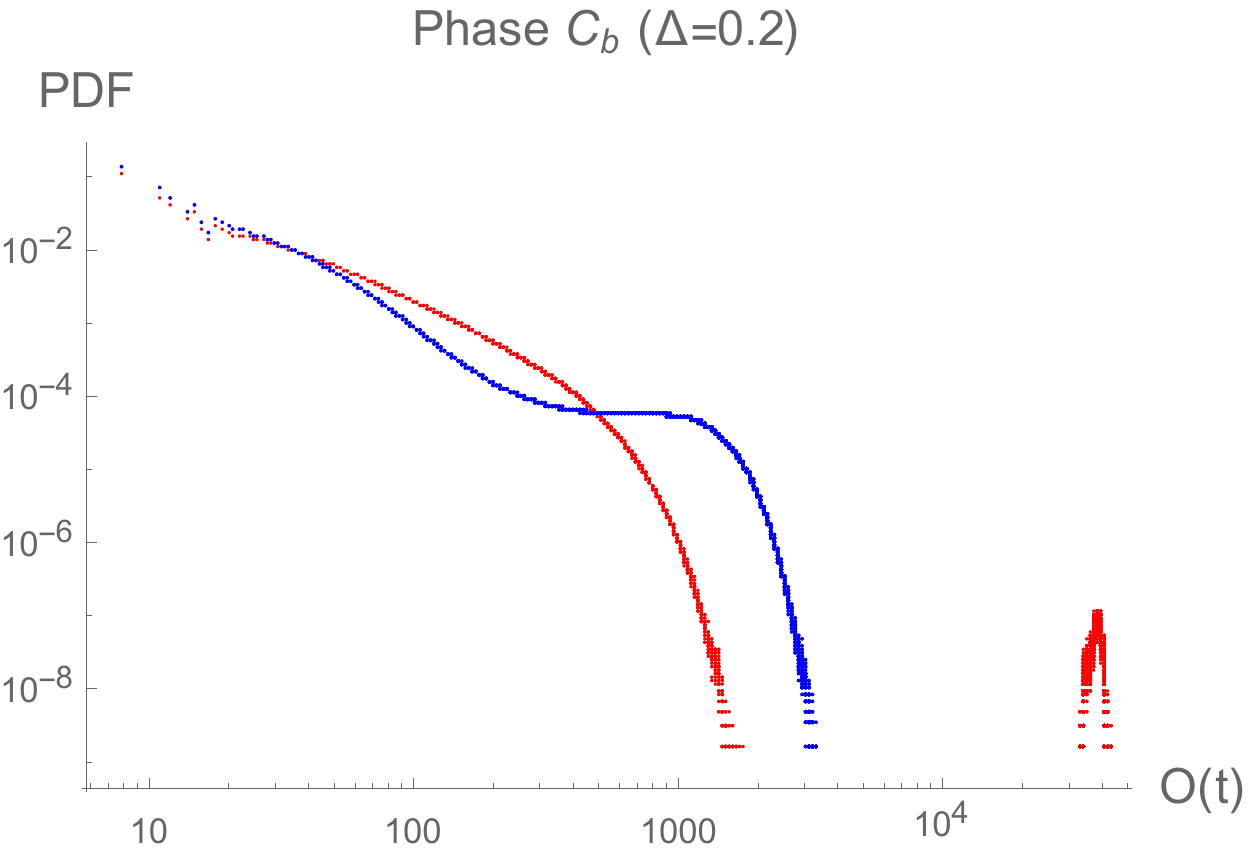}
\caption[aaa]{\small {Histograms of vertex coordination numbers measured in the de Sitter phase $C_{dS}$ (left) and in the bifurcation phase $C_b$ (right).  Blue data points are for the central slice $t_c$  
and  red data points are for $t_c + 1$ (see footnote \footnote{}).
In phase $C_{dS}$ there is no clear difference between the two distributions whereas in   phase $C_b$ the distributions look very different. The difference is due to a single highest order vertex present in $t_c\pm 1$ (and also in $t_c\pm 3$, $t_c\pm 5$, ...), which is not present in $t_c$ (nor in $t_c\pm 2$, $t_c\pm 4$, ...). The data was averaged over individual triangulations after performing the centering procedure described in footnote $^3$. The highest order vertex coordination number observed in phase $C_b$ has an  approximately Gaussian distribution centred  around $10-100$ times the coordination number of other high order vertices present in $t_c + 1$. The result is that in phase $C_b$ one observes a clear gap in the coordination number histograms  in odd $t$, whereas there is no such gap in even $t$. No gap is visible in phase $C_{dS}$.}}
\label{nocoord}
\end{figure}
\footnotetext{{The central time coordinate $t_c$ is defined as a lattice time for which the maximal coordination number of a vertex  $O_{max}(t)$ is the most symmetric with respect to $|t-t_c|$ and additionally it is assumed that the highest order vertex in the whole triangulation $O_{max}(t_0)$ is placed in odd $t$ (division into odd and even  time slices is compatible with the observed properties of phase $C_b$).  $t_c$ performs a slow random walk around the periodic time axis. In averaging over triangulations one gets rid of this translational zero mode by redefining the time coordinate such that for each triangulation  $t_c = 40$. }}

\begin{figure}[H]
\centering
\includegraphics[width=0.45\linewidth]{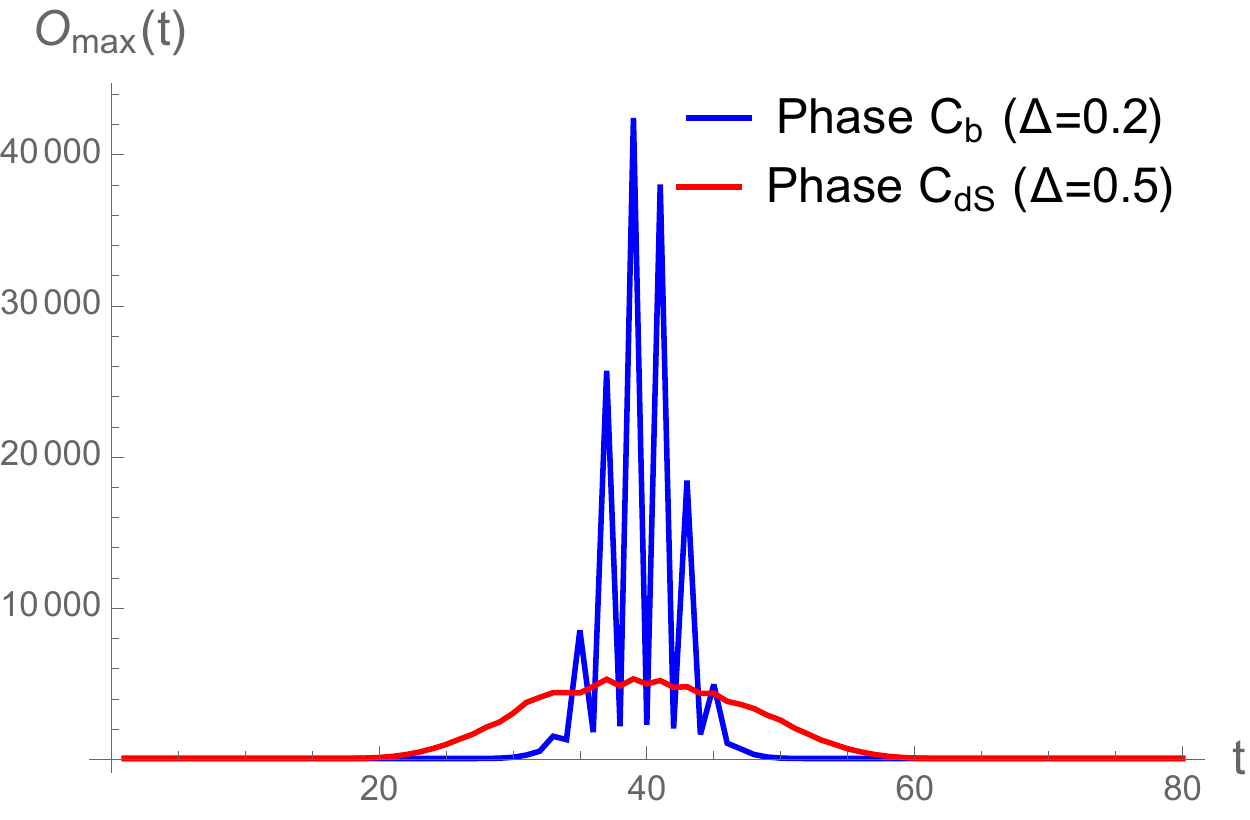}
\includegraphics[width=0.45\linewidth]{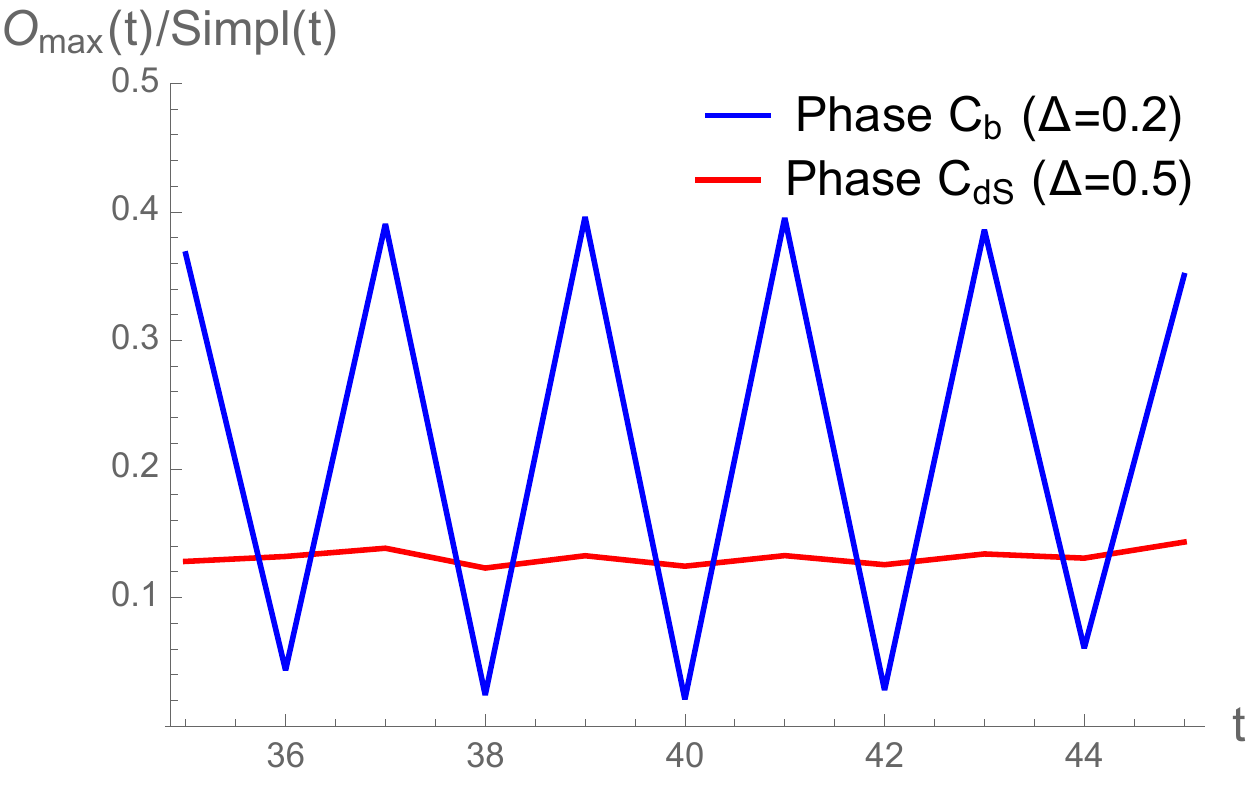}
\caption{\small {Left: The maximal coordination number of a vertex $O_{max}(t)$ plotted as a function of lattice time coordinate $t$. The data were averaged over individual triangulations after performing the centering procedure described in footnote $^3$. Right: The same chart after normalising the maximal coordination number by dividing $O_{max}(t)$ by the total number of four-simplices having at least one vertex at time $t$ (only a central part of the triangulation: $t_c\pm 5$ is shown which is consistent with the extended  part of the CDT universe). In the bifurcation phase $C_b$ the maximal coordination number jumps between odd and even time slices. There is no such jumping in the de Sitter phase $C_{dS}$.}}
\label{Omax}
\end{figure}

When one looks inside the de Sitter phase $C_{dS}$ and measures  the distribution of vertex coordination numbers\footnote{Vertex coordination number is defined as a number of 4-simplices  sharing a given vertex.}  (see Fig. \ref{nocoord} left) one observes that there is no clear gap between the highest order vertex $O_{max}\big(t\big) $ (the one with maximal coordination number in a given time slice $t$) and other vertices present in the same slice $O\big(t\big) $. This is independent on the parity of the lattice time coordinate $t$. The situation changes when one goes inside the bifurcation phase $C_b$. Here the distribution of vertex coordination numbers (see Fig. \ref{nocoord}  right)  depends on $t$. For (say) even $t$  the distribution is quite similar to the one observed in phase $C_{dS}$ whereas for odd $t$ one observes a clear gap between the highest order vertex    $O_{max}\big(t\big) $ and other vertices $O\big(t\big) $. The gap rises when one goes deeper and deeper into the bifurcation phase and also when one increases the total  lattice volume $N_{4,1}$. As a result the maximal coordination number $O_{max}\big(t\big) $  {jumps}  between odd and even spatial slices in the bifurcation phase $C_b$, and there is no such {jumping} in phase $C_{dS}$ {(see Fig. \ref{Omax})}.
In Refs. \cite{Ambjorn:2015qja,   Coumbe:2015oaa} a simple order parameter based on the above observation was proposed
\begin{equation}\label{op}
\text{OP}_2= \frac{1}{2} \left[ \Big| O_{max}\big(t_0\big) - O_{max}\big(t_0+1\big) \Big|   +     \Big| O_{max}\big(t_0\big) - O_{max}\big(t_0-1\big) \Big|\right],
\end{equation}
where $t_0$ is chosen in such a way that $ O_{max}\big(t_0\big)$ is  the highest coordination number in a triangulation, i.e. the highest among all high order vertices $ O_{max}\big(t\big)$:
\begin{equation}
O_{max}\big(t_0\big) = \max_t  O_{max}\big(t\big) \ .
\end{equation}

\begin{figure}[H]
\centering
\includegraphics[width=0.45\linewidth]{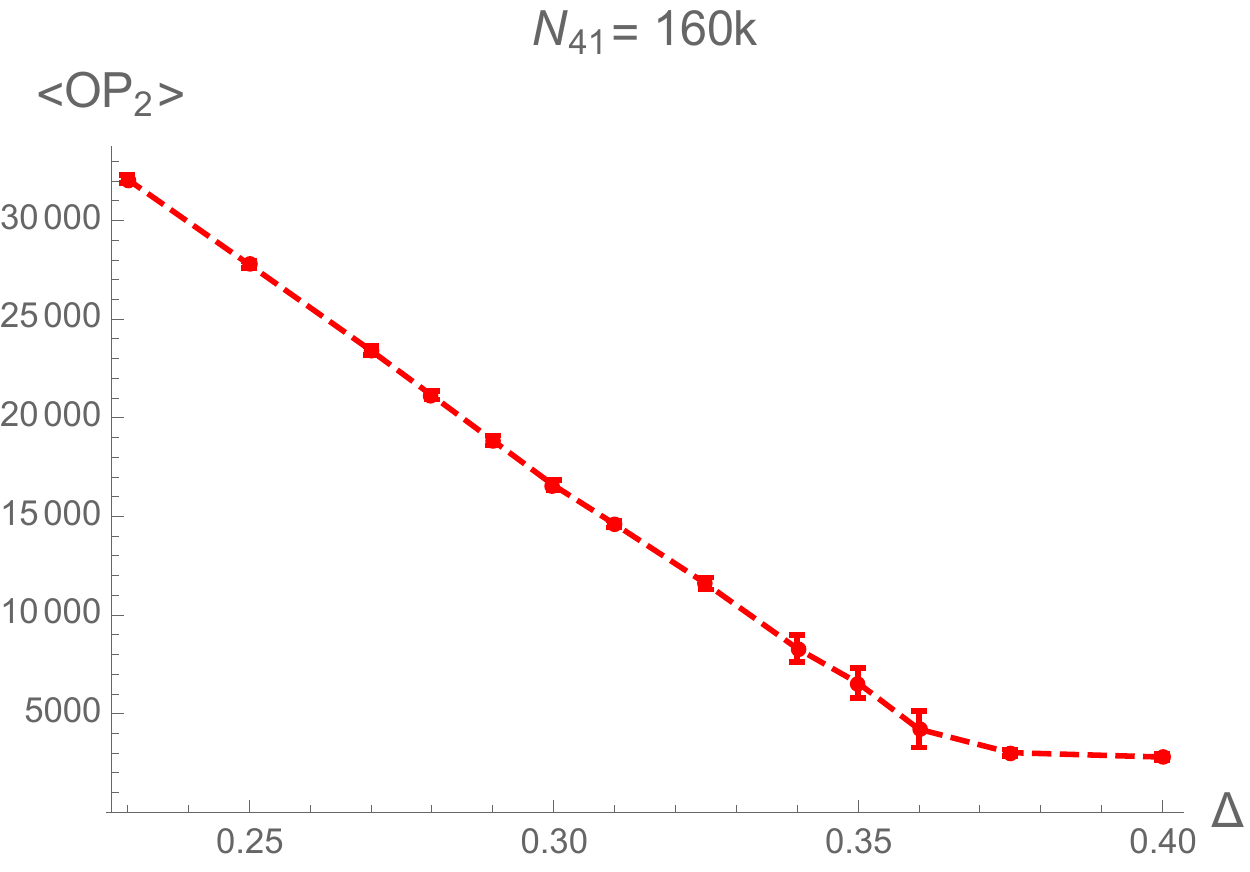}
\includegraphics[width=0.45\linewidth]{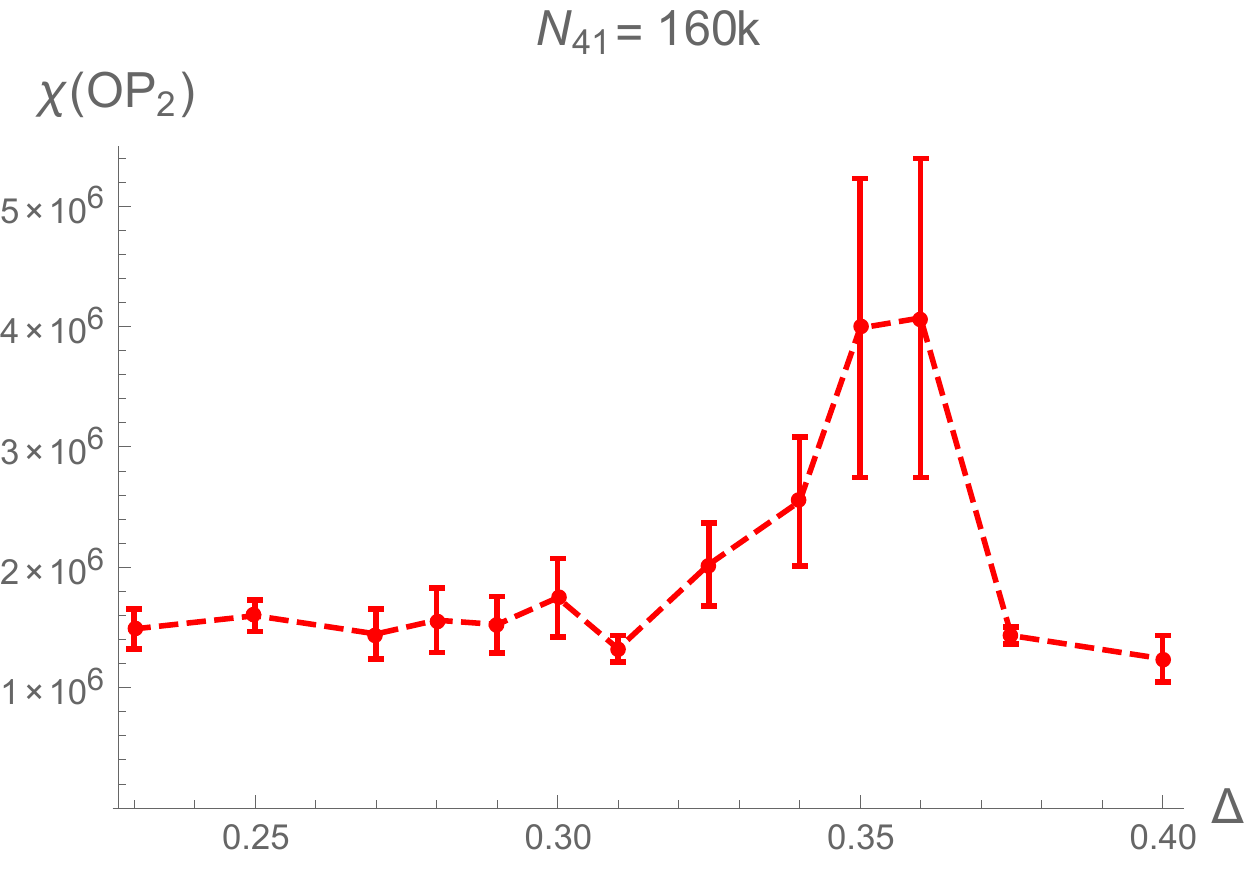}
\caption{\small The order parameter mean value $\langle \text{OP}_2 \rangle$ (left) and its susceptibility $\chi_{\text{OP}_2}$ (right) as a function of $\Delta$.}
\label{OP2fig}
\end{figure}

The order parameter is approximately zero in the (more) symmetric phase $C_{dS}$ and non-zero in the symmetry-broken phase $C_b$. If, for example, one starts from some chosen point in the phase diagram $(\kappa_0, \Delta)$ inside phase $C_{dS}$ and lowers $\Delta$ one  encounters the phase transition to phase $C_b$ when the order parameter starts to rise  approximately linearly with decreasing $\Delta$ (see Fig. \ref{OP2fig} left).
The (pseudo-)critical point $\Delta^{crit}$ is signalled by a peak in  susceptibility (Fig. \ref{OP2fig} right) and, as already explained, its position 
depends on the lattice volume $N_{4,1}$. One can fit the measured volume dependence to the formula
\begin{equation}\label{voldep}
\Delta^{crit}\big(N_{4,1}\big)=\Delta^{crit}\big(\infty\big) - \alpha \ N_{4,1}^{\ -1/\gamma}
\end{equation}
and compute the critical exponent $\gamma$. A first order transition should be associated with  $\gamma = 1$, and accordingly  $\gamma \neq 1$ signals a higher order transition. 
In Fig. \ref{VolDep} we present results obtained for a wider choice of lattice volumes  $N_{4,1}$ and also much longer Monte Carlo runs than in reference \cite{Coumbe:2015oaa}. The critical exponent  fitted using formula (\ref{voldep}) is $\gamma = 2.71 \pm 0.34$ which is greater than $1$ with a confidence interval of $99 \%$. This result strongly supports the conjecture that the $C_{dS}-C_b$ phase transition is a higher order transition. For comparison we also present in Fig. \ref{VolDep} a fit with forced value of critical exponent $\gamma=1$ which seems to be much  less likely.

\begin{figure}[H]
\centering
\includegraphics[width=0.55\linewidth]{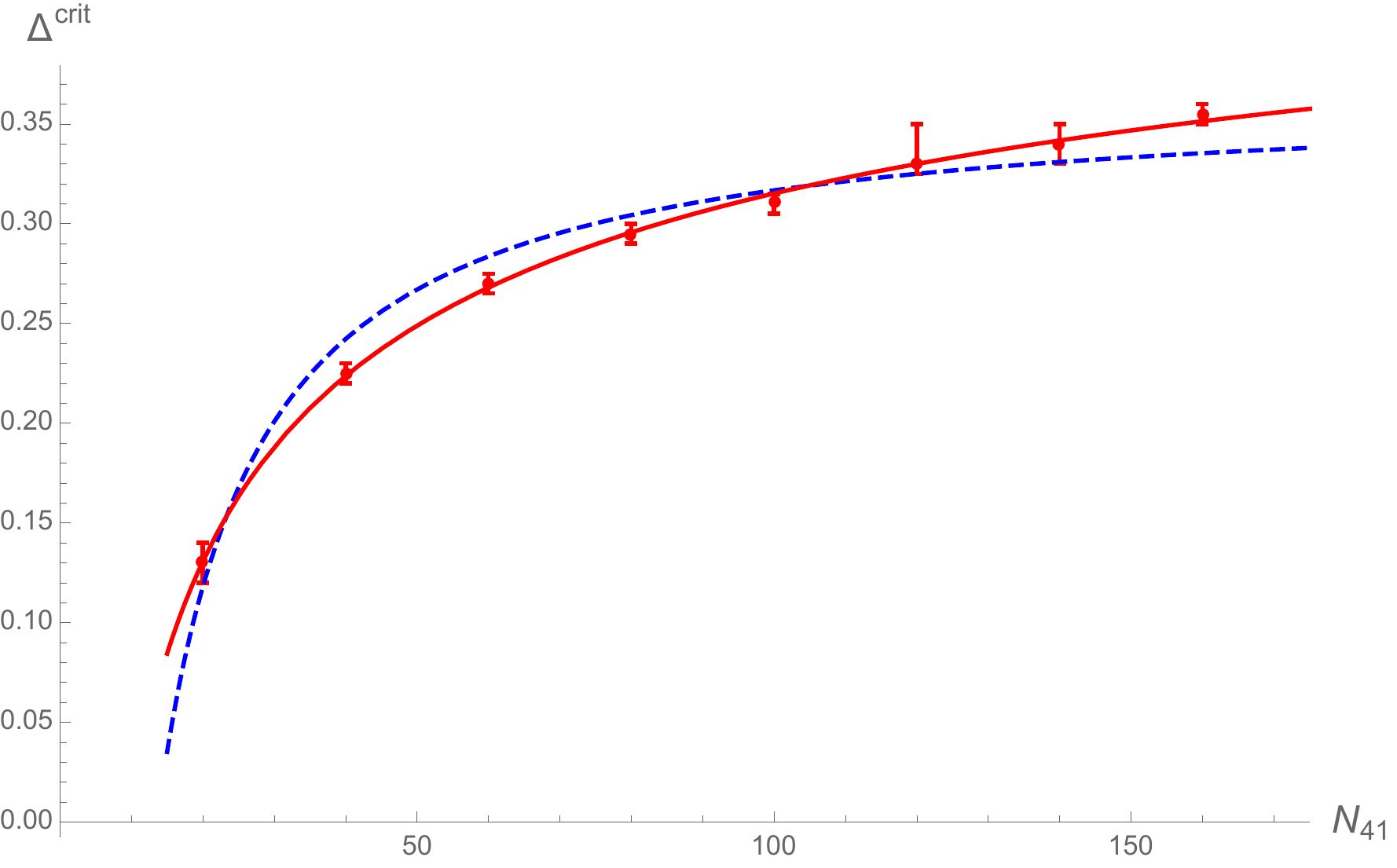}
\caption{\small Lattice volume dependence of (pseudo-)critical points $\Delta^{crit}$ (bars) together with a fit of formula (\ref{voldep})  (red line) and the same fit  with a forced value of the critical exponent $\gamma=1$ (blue dashed line). }
\label{VolDep}
\end{figure}

A practical problem with using formula (\ref{voldep}), and the reason why we present the updated data,  is that one should have transition points measured for a wide choice of  lattice volumes, which is very computationally expensive. This is due to the fact that when performing numerical simulations near phase transitions of higher order  one usually encounters a so-called critical slowing down, related to very large auto-correlation times of the measured data. The auto-correlation time (in Monte Carlo time) peaks at the phase transition  causing the numerical algorithm to loose efficiency, and consequently a very long simulation time is needed to get reliable data in the vicinity of the transition point. This kind of critical slowdown is  clearly observed for the $C_{dS}-C_{b}$  phase transition - see Fig \ref{Corelfig} (left) where the measured auto-correlation of the $\text{OP}_2$ order parameter 
\begin{equation}\label{autocor}
\text{AC}_{\text{OP}_2} (\Delta \tau)= \frac {\big \langle \text{OP}_2 (\tau)   \text{OP}_2 (\tau+\Delta \tau)\big \rangle_\tau - \big \langle \text{OP}_2 (\tau)  \big \rangle_\tau  \big \langle  \text{OP}_2 (\tau+\Delta \tau)\big \rangle_\tau  } {\big \langle \text{OP}_2^{\ 2} (\tau) \big \rangle_\tau- \big \langle \text{OP}_2 (\tau) \big \rangle_\tau^{2}}
\end{equation}
is shown as a function of the Monte Carlo time difference $\Delta \tau$.
Red data points present auto-correlation at the phase transition  while other colours are auto-correlations observed slightly away from the phase transition point. One clearly sees that  auto-correlation is much longer in the vicinity of the phase transition. This difference can be also exploited to signal the position of (pseudo-)critical points. Fig. \ref{Corelfig} (right) presents the auto-correlation time around phase transition points measured for various lattice volumes. The transition points $\Delta^{crit}$ are defined by  peaks in susceptibility (\ref{susceptibility}) and the auto-correlation time $\tau_{ac}$ is obtained by fitting
\begin{equation}\label{autotime}
\text{AC}_{\text{OP}_2} (\Delta \tau)= {\cal N} \exp{\big( - \Delta \tau / \tau_{ac}\big )}
\end{equation}
to the empirical auto-correlation data (\ref{autocor}). One clearly sees that the peaks in auto-correlation time are consistent with the peaks in susceptibility. 
As a side effect the very long auto-correlation time at the phase transition means that a much longer simulation time\footnote{Monte Carlo simulations needed to produce susceptibility plots with reasonable error bars and as a consequence to produce Fig. \ref{VolDep} lasted almost one year.} is needed to decrease the error bars\footnote{Measurement errors were estimated  using a
single-elimination (binned) jackknife procedure, after blocking the data to account for
auto-correlation errors. The procedure was described in detail in Ref. \cite{ Coumbe:2015oaa}.} of the measured observables in the vicinity of the phase transition points. This explains the relatively large error bars observed for the  transition points, e.g. in Fig. \ref{OP2fig}.

\begin{figure}[h]
\centering
\includegraphics[width=0.45\linewidth]{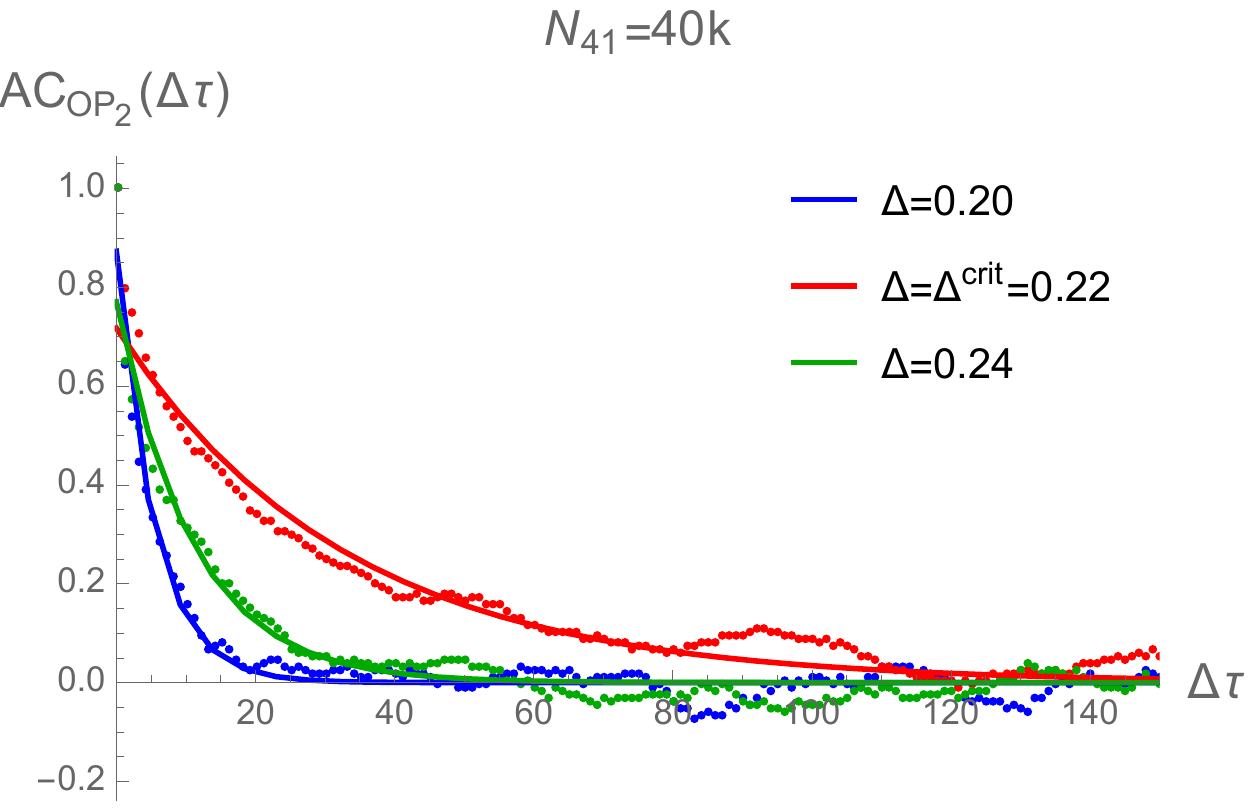}
\includegraphics[width=0.45\linewidth]{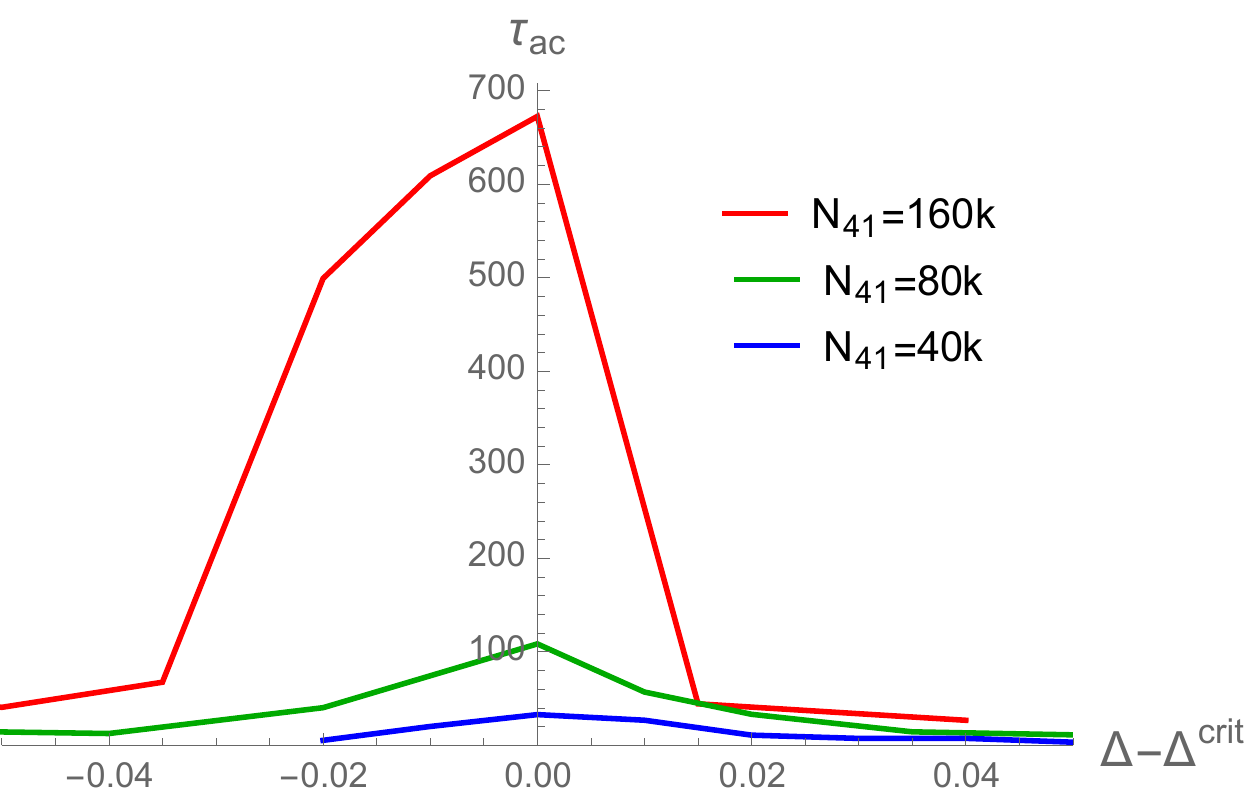}
\caption{\small Left: Auto-correlation of the measured order parameter $\text{OP}_2$ calculated according to Eq. (\ref{autocor}) for the transition point (red points) and slightly away from the transition point (blue and green points) and the fits of formula (\ref{autotime}) to the measured data. The unit of the horizontal axis is $10^9$ attempted Monte Carlo moves. Right: Auto-correlation time $\tau_{ac}$ obtained by fitting formula (\ref{autotime}) to the measured $\text{OP}_2$ auto-correlation data plotted as a function of $\Delta - \Delta^{crit}$ for various lattice volumes. The values of $ \Delta^{crit}$ were established for each lattice volume separately by looking at the peaks of susceptibility $\chi_{\text{OP}_2}$. Peaks in auto-correlation time are consistent with peaks in susceptibility.}
\label{Corelfig}
\end{figure}

Last but not least, we comment on the double peaks observed in the $\text{OP}_2$ order parameter histograms  (see Fig \ref{Histograms} left) measured at the $C_{dS}-C_b$ phase transition points ($\Delta \approx \Delta^{crit}$) as already reported in Ref. \cite{ Coumbe:2015oaa}. The double peaks are caused by the order parameter jumping (in Monte Carlo time) between two values (see Fig. \ref{OPfig}) and  it   was  noticed that the frequency of such jumps decreases with increasing lattice volume (see Fig. \ref{OPfig} left), which might in principle signal a first order transition. 
Now we attribute the decrease in jumping frequency or, in other words, an increase in  jumping period to the auto-correlation time which also increases with increasing lattice volume. If one, for example, introduces a dimensionless  simulation time by rescaling $\tau \to \tau/ \tau_{ac}$ to account for the auto-correlation difference observed for various lattice volumes $N_{4,1}$ one can see that the jumps are less frequent for smaller lattice volumes than they are for larger lattice volumes (see Fig. \ref{OPfig} right). 
One can also argue that one should look at the normalised order parameters 
to account for volume difference.  Vertex coordination number  scales approximately linearly with the number of simplices in a triangulation and thus one should look at $\text{OP}_2 / N_{4,1}$  rather than $\text{OP}_2$. Consequently the amplitude of the (normalised)  order parameter jumps seems to decrease with increasing lattice volume. It is clearly visible if one fits a double Gaussian function to the measured   histograms of $\text{OP}_2 / N_{4,1}$ (see Fig. Fig. \ref{Histograms} left). One observes that the Gaussians  are only slightly separated. In Fig. \ref{Histograms} (right) we plot the separation of the two peaks as a function of  lattice volume $N_{4,1}$. The separation seems to decrease with increasing lattice volume but 
we have not yet reached a point at which it shrinks to zero. This analysis suggests that the spurious  behaviour of $\text{OP}_2$ which mimics some features of a first order transition is  most likely due to finite size effects, and it supports the conjecture that the $C_{dS}-C_b$ phase transition is really a higher order transition. Similar phenomena were previously observed for the transition between the bifurcation phase $C_b$ and phase $B$ (formerly called the $C-B$  transition), which was shown to be a higher order transition \cite{Ambjorn:2012ij}, and it was recently explained by a very nontrivial shape of free energy in the vicinity of the phase transition line \cite{Ambjorn:2016mnn}.

\begin{figure}[h]
\centering
\includegraphics[width=0.45\linewidth]{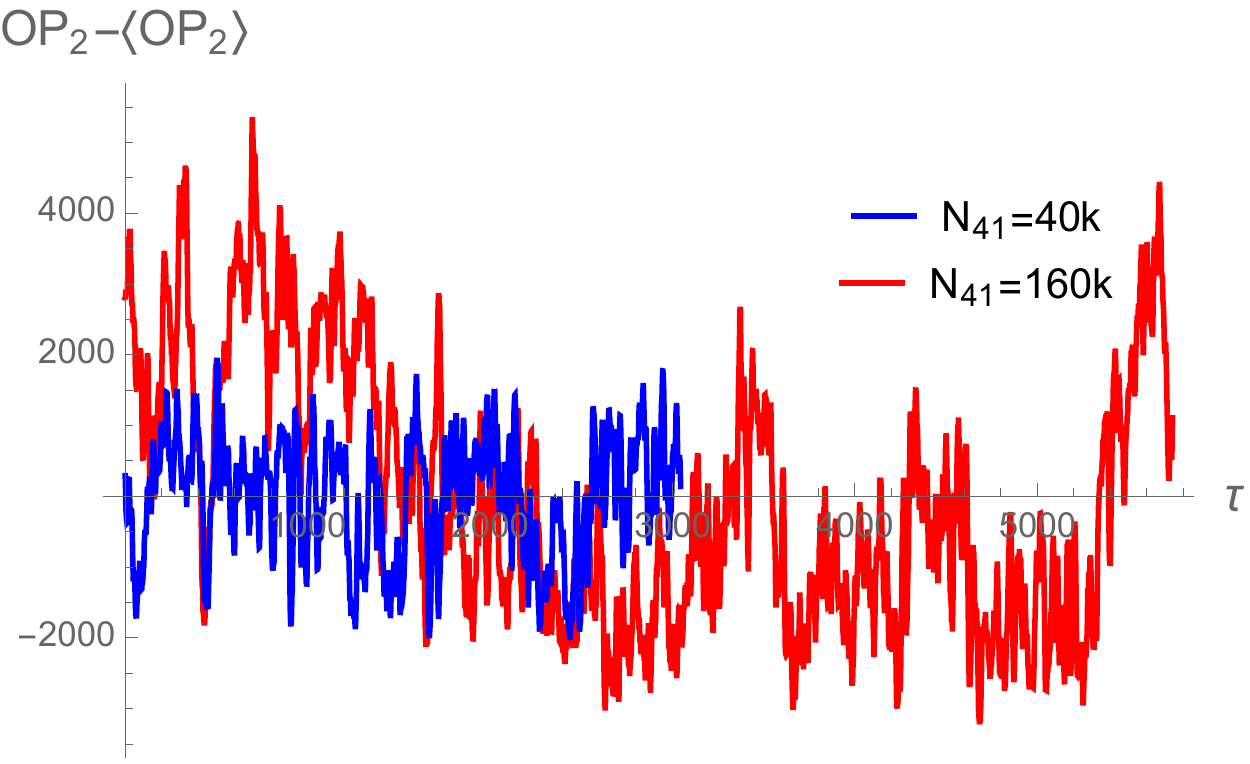}
\includegraphics[width=0.45\linewidth]{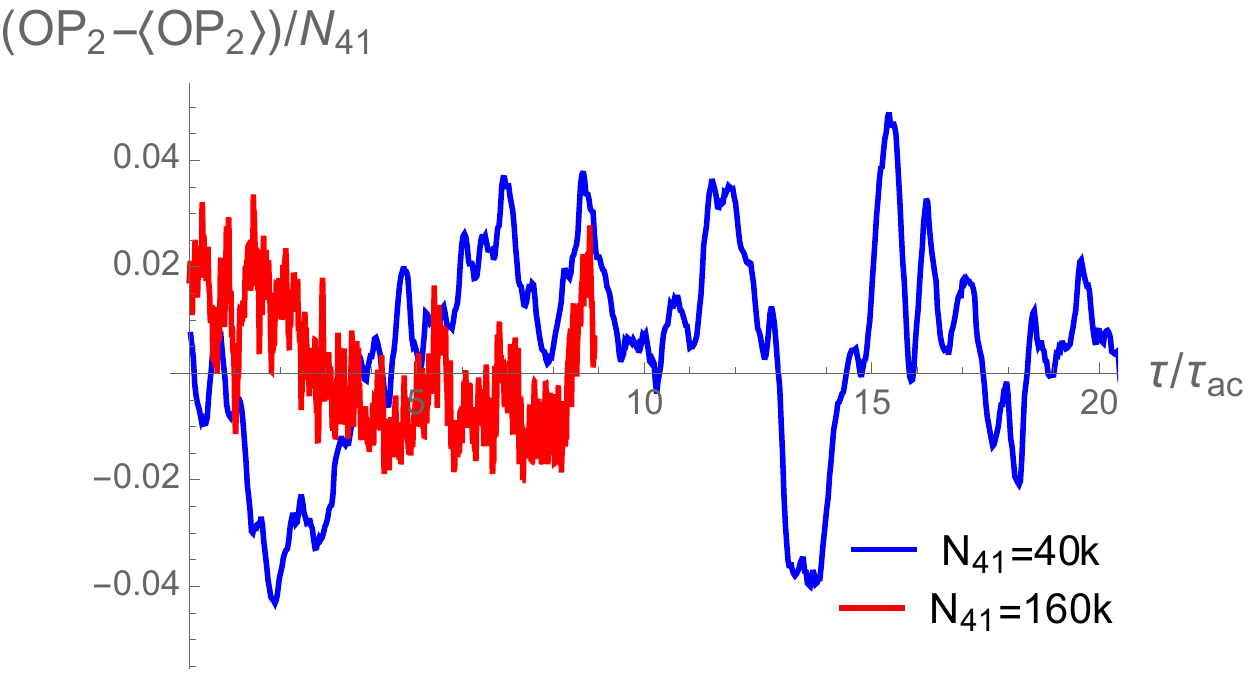}
\caption{\small Left: The order parameter $\text{OP}_2$ plotted as a function of Monte Carlo time (the unit of the horizontal axis is $10^9$ attempted moves). In order to compare data measured for different lattice volumes we use $\text{OP}_2 - \langle \text{OP}_2 \rangle$ and to
 smooth out small oscillations the data was averaged over $100$ consecutive values (moving average). The order parameter jumps between two levels. The amplitude of jumps and the jumping period increase with increasing lattice volume. Right: the same data of the order parameter after a rescaling of Monte Carlo time $\tau \to \tau / \tau_{ac}$, where $\tau_{ac}$  is the auto-correlation time defined by formula (\ref{autotime}), and $\text{OP}_2 \to \text{OP}_2 / N_{4,1}$. In this scenario both the amplitude of jumps and the jumping period decrease with increasing lattice volume. }
\label{OPfig}
\end{figure}

\begin{figure}[H]
\centering
\includegraphics[width=0.45\linewidth]{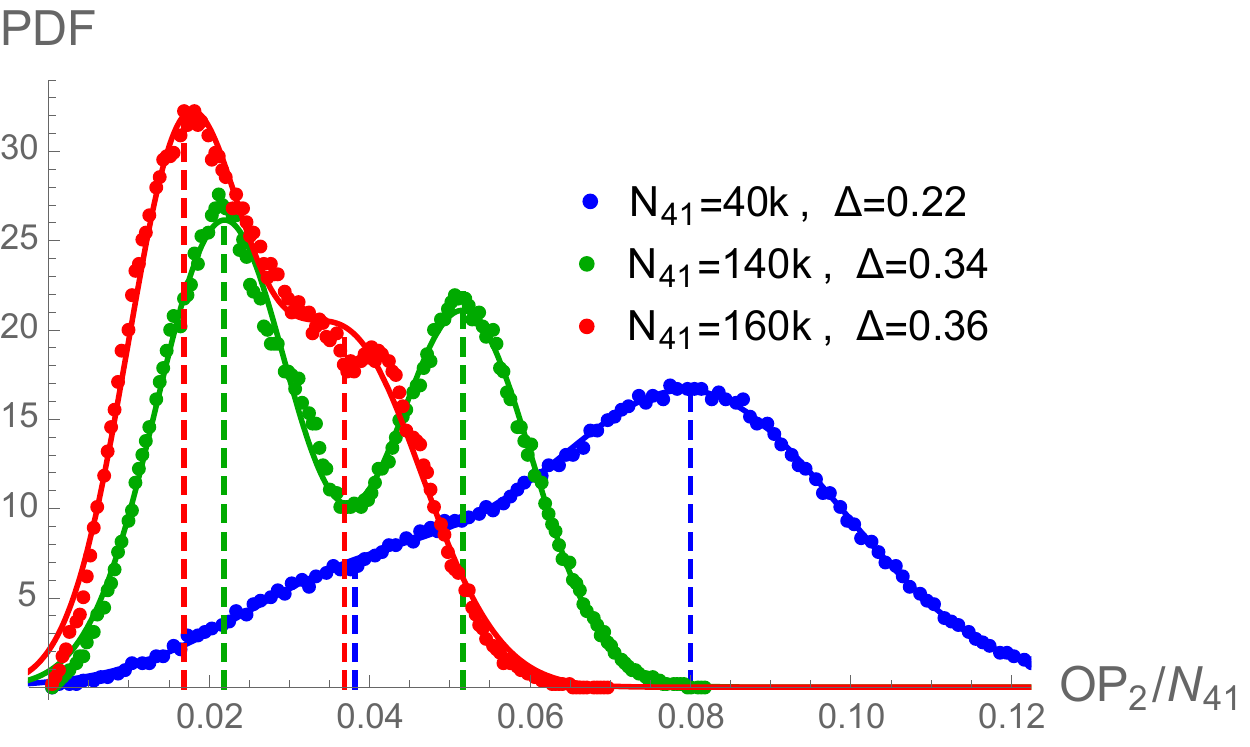}
\includegraphics[width=0.45\linewidth]{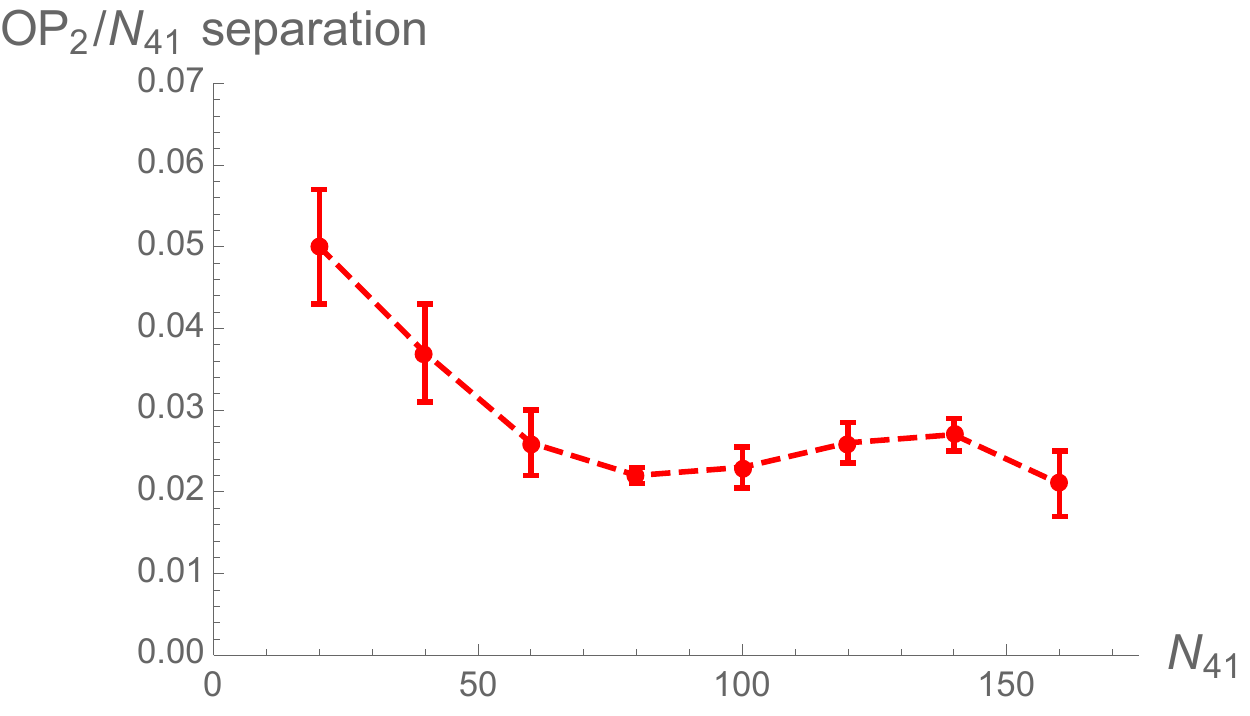}
\caption{\small Left: Histograms of the (normalised) order parameter $\text{OP}_2/ N_{4,1}$ measured at  (pseudo-)critical  points ($\Delta = \Delta^{crit}$) defined by the peaks in susceptibility $\chi_{\text{OP}_2}$ for various lattice volumes. One can observe the double peaks related to the order parameter jumping between two states. The double peak structure is clearly visible for $N_{4,1} = 140$k where the height of the two peaks is (almost) the same. The double peaks are slightly less visible for $N_{4,1} = 40 $k and $N_{4,1} = 160$k where the height of the peaks is different. This is due to the fact that for  $N_{4,1} = 140 k$ the data were measured for $\Delta^{crit}$ fixed very precisely at the (pseudo-)critical point while for other volumes $\Delta^{crit}$ was  set slightly away from the true (pseudo-)critical point. The chart also shows the fits of the double Gaussian functions to the measured data (lines). The positions of the two peaks (from the fits) are marked by dashed lines. Right: Separation of the two peaks in $\text{OP}_2/ N_{4,1}$  histograms calculated from the double Gaussian fits plotted as a function of  lattice volume. }
\label{Histograms}
\end{figure}

\end{section}

\begin{section}{Adding $N$ massless scalar fields to CDT}\label{fields}






Motivated by the suggestion of Hartle and Hawking that a sufficient number of matter fields may be a necessary condition to produce the correct classical behaviour of the universe \cite{Hartle:2008ng} we investigate the effect of adding $N$ massless scalar fields to the bare lattice action of CDT. 
We discuss the impact of the scalar fields on average spatial volume profiles and spatial volume fluctuations in the de Sitter phase $C_{dS}$ and the bifurcation phase $C_b$. 
We also analyse whether the position of the $C_{dS}-C_b $ transition line is dependent on the number of massless scalar fields $N$. 

To this end, we employ a bare action of the form $S(T,x) = S_{EH} (T) + S_M(T,x)$, where $S_{EH} (T)$ is a bare CDT action for pure gravity (\ref{eq:GeneralEinstein-ReggeAction}) and $S_M(T,x)$ is the action for $N$ copies of minimally coupled  scalar fields $x$,

\begin{equation}\label{actionFields}
S_{M}(T,x)=\frac{1}{2}\sum_{F=1}^{N}\sum_{i}{\mu^{2}}\left(x_{i}^{F}\right)^{2}+\frac{1}{2}\sum_{F=1}^{N}\sum_{i\leftrightarrow j}\left(x_{i}^{F}-x_{j}^{F}\right)^{2},
\end{equation}

\noindent with a measure

\begin{equation}
\mathcal{D}[x]=\prod_{F=1}^N\prod_{i}\frac{dx_{i}^{F}}{\sqrt{\pi}}.
\end{equation}

\noindent In (\ref{actionFields}) we assume that the (real valued) scalar fields are located in simplex centres, and we  express the scalar field action in terms of the dual lattice. Consequently, 
 the  sums $\sum_{i}$ and $\sum_{i\leftrightarrow j}$ are over all 4-simplices and over all neighbouring pairs of simplices, respectively. In this work we are only interested in massless scalar fields and so we set the mass parameter $\mu$ equal to zero.

In order to generate configurations, which now consist of a triangulation and superimposed scalar fields,
according to action (\ref{actionFields}), we modify the  Metropolis algorithm used thus far.
The heat bath method is applied to update the values of the scalar fields.
Incorporating the scalar fields does not change the geometrical structure of the Monte Carlo moves, 
but it influences their weight so that the detailed balance condition is fulfilled \cite{Ambjorn:2012jv}.
Due to the quadratic form of the scalar field action $S_M[T, x]$ (\ref{actionFields}),
the heat bath method reduces to generating 
the updated values of the scalar fields inside the region in which the moves are implemented, i.e. simplices affected by a given move, from a multivariate Gaussian distribution
whose parameters depend on the surrounding field values.
The scalar fields are updated regardless of whether the move is accepted or not.
Such a method is very efficient as
the field values inside the region in which the moves are implemented are always altered and 
the acceptance rate is not impaired.
The factors which are not covered by the Gaussian distribution
depend only on the field values around the move region
and contribute to the weight of the move.
The normalisation factors (matrix determinants) 
can be absorbed by bare coupling redefinitions.

\begin{figure}[h]
\centering
\includegraphics[width=0.45\linewidth]{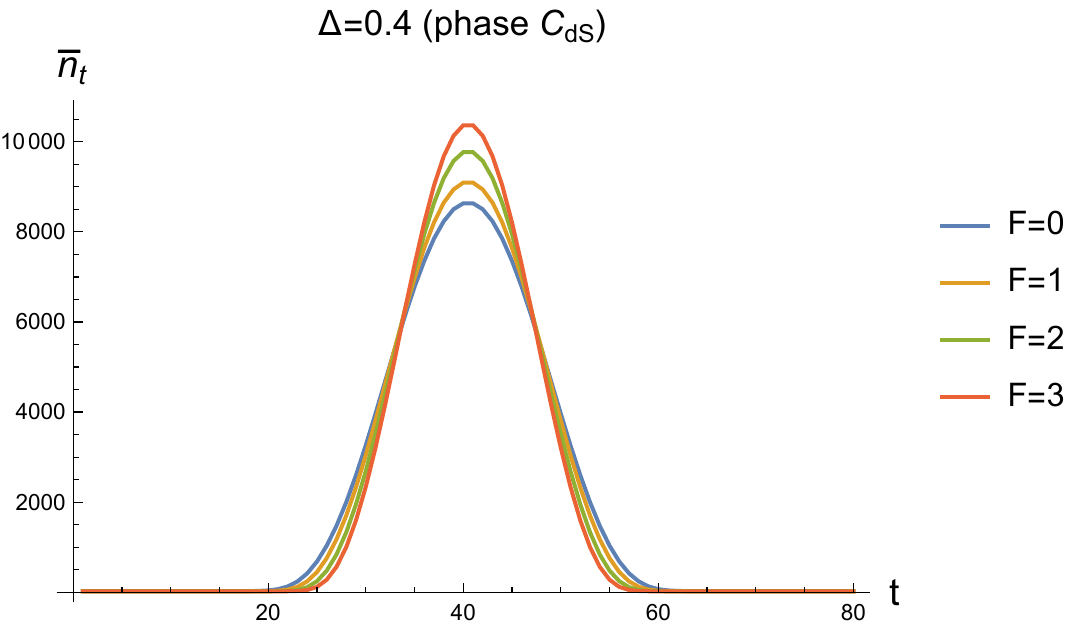}
\includegraphics[width=0.45\linewidth]{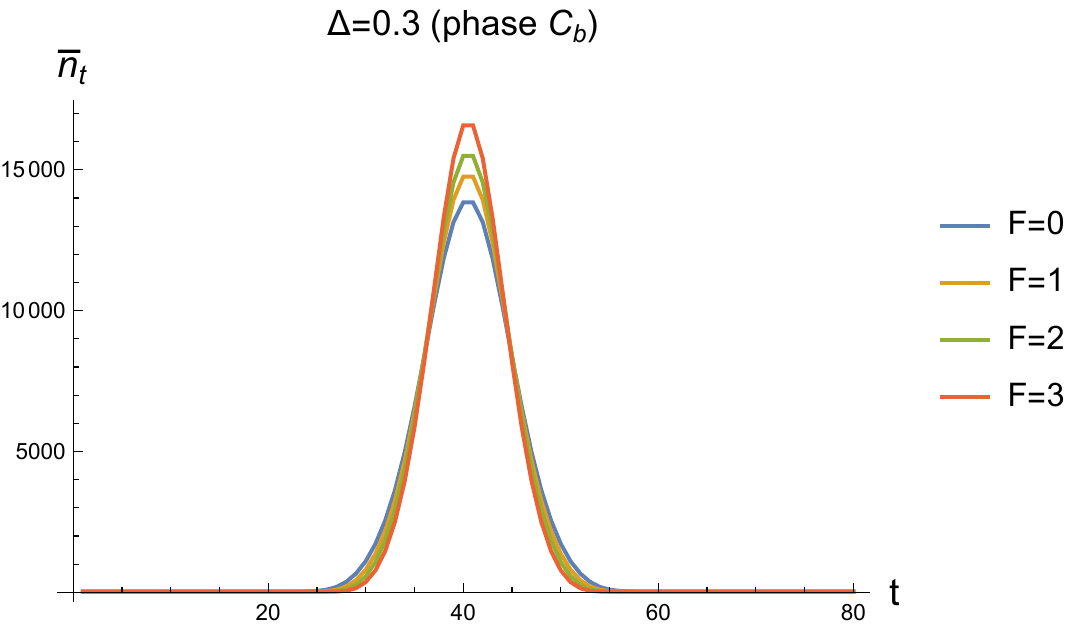}
\caption{\small Spatial volume profiles  for CDT including 0, 1, 2 and 3 massless scalar fields.}
\label{avfig}
\end{figure}

\begin{figure}[h]
\centering
\includegraphics[width=0.45\linewidth]{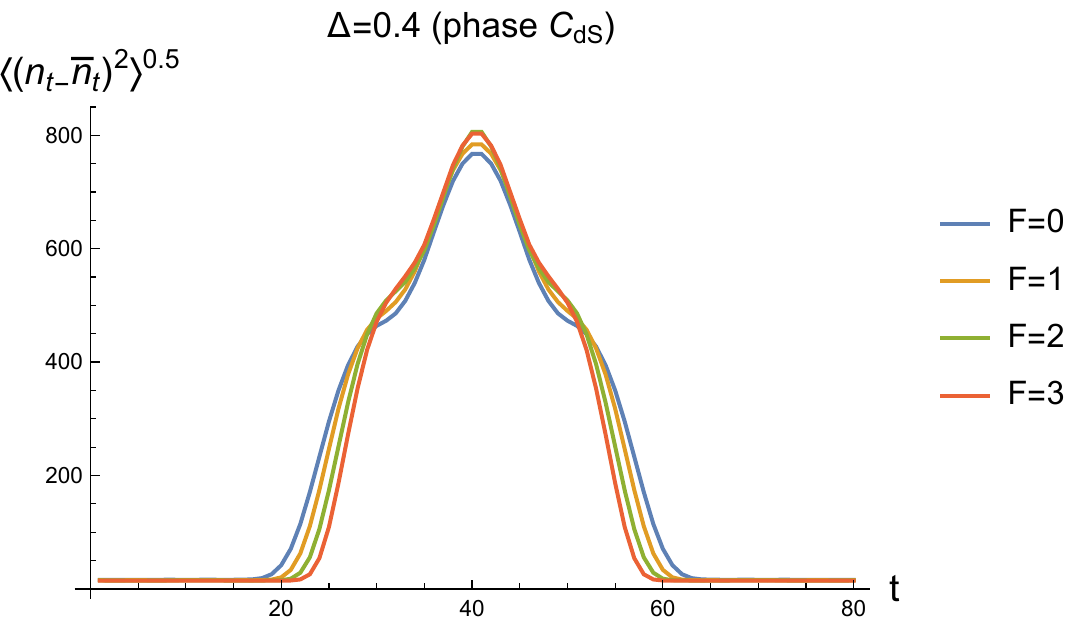}
\includegraphics[width=0.45\linewidth]{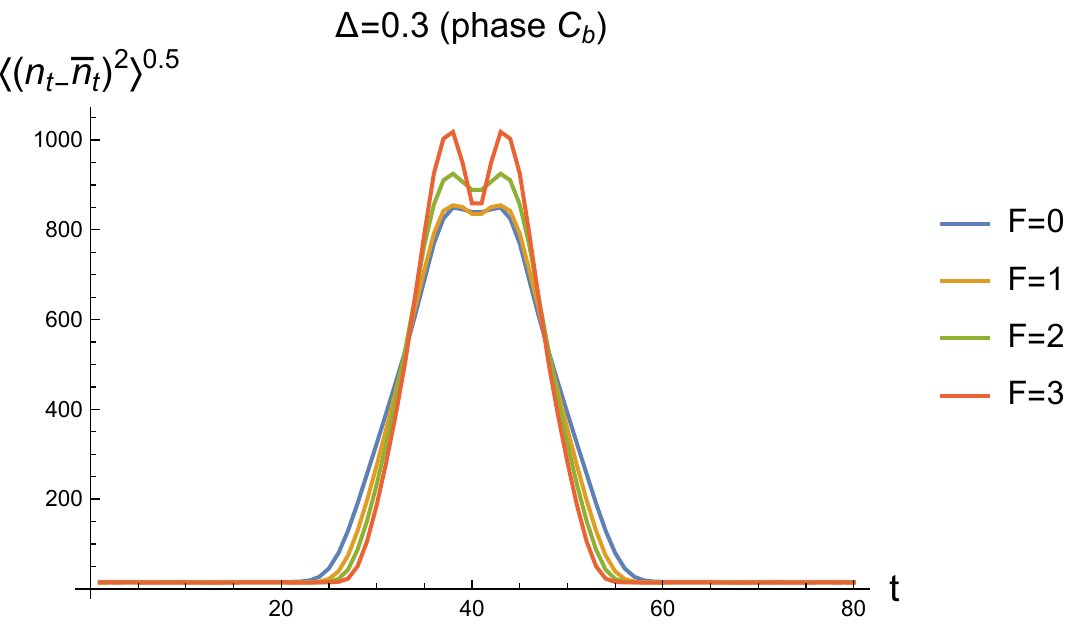}
\caption{\small Fluctuations of spatial volume profiles  for CDT including 0, 1, 2 and 3 massless scalar fields.}
\label{sdfig}
\end{figure}

A question that arises is how do the scalar fields impact the spacetime geometry. Here we concentrate on the (average) spatial volume profiles $ \bar n_t  \equiv \langle N_{4,1}(t) \rangle$ (see Fig. \ref{avfig}) and fluctuation amplitudes 
$ \big\langle \big (N_{4,1}(t)  - \bar n_t \big)^2 \big \rangle^{1/2} $ (see Fig. \ref{sdfig}), where the averages $\langle .\rangle$ are taken over lattice configurations. The data  measured in  the de Sitter phase $C_{dS}$ are shown on the left and in the bifurcation phase $C_b$ on the right charts, respectively. As a result of adding scalar fields both the volume and fluctuation profiles narrow in the time direction. The effect is qualitatively the same in the de Sitter phase $C_{dS}$ and in the bifurcation phase $C_b$. Additionally, in phase $C_b$ one may observe that adding scalar fields leads to a greater decrease in volume fluctuations in the centre of the profile.

Using the $\text{OP}_2$ order parameter defined in Eq. (\ref{op})  we also analyse the impact of the scalar fields on the position of the $C_{dS}-C_b $ phase transition line. Once again we fix  $\kappa_0$ and  vary $\Delta$ to find the transition point  $\Delta^{crit}$ by looking for peaks in the $\text{OP}_2$  susceptibility $\chi_{\text{OP}_2}$ (\ref{susceptibility}).
Figure~\ref{FigFields1} shows the mean values of the order parameter $\langle \text{OP}_2 \rangle$ and susceptibility $\chi_{\text{OP}_2}$ as  functions of $\Delta$ for CDT including 1, 2 and 3 massless scalar fields. The peak value of $\chi_{\text{OP}_2}$ indicates that the (pseudo)-critical value of $\Delta$ seems to be largely independent of the number of massless scalar fields, remaining within the range $\Delta^{crit}\approx 0.36$ - $0.38$, which is very close to  $\Delta^{crit}=0.36\pm0.01$ observed for pure gravity simulations with the same lattice volume ($N_{4,1}=160000$). However, the plots of $\langle \text{OP}_2 \rangle$ may suggest that the (pseudo)-critical value of $\Delta$ slightly increases  in response to an increasing number of scalar fields, although this is far from conclusive given the data presented. To summarise, the data presented in Fig.~\ref{FigFields1} indicates that adding $N$ massless scalar fields to CDT does not significantly alter the position of the $C_{dS}-C_{b}$ transition, suggesting that the bifurcation phase is probably not simply an artifact of the naive pure gravity formulation of CDT.

\begin{figure}[H]
  \centering
  \scalebox{.6}{\includegraphics{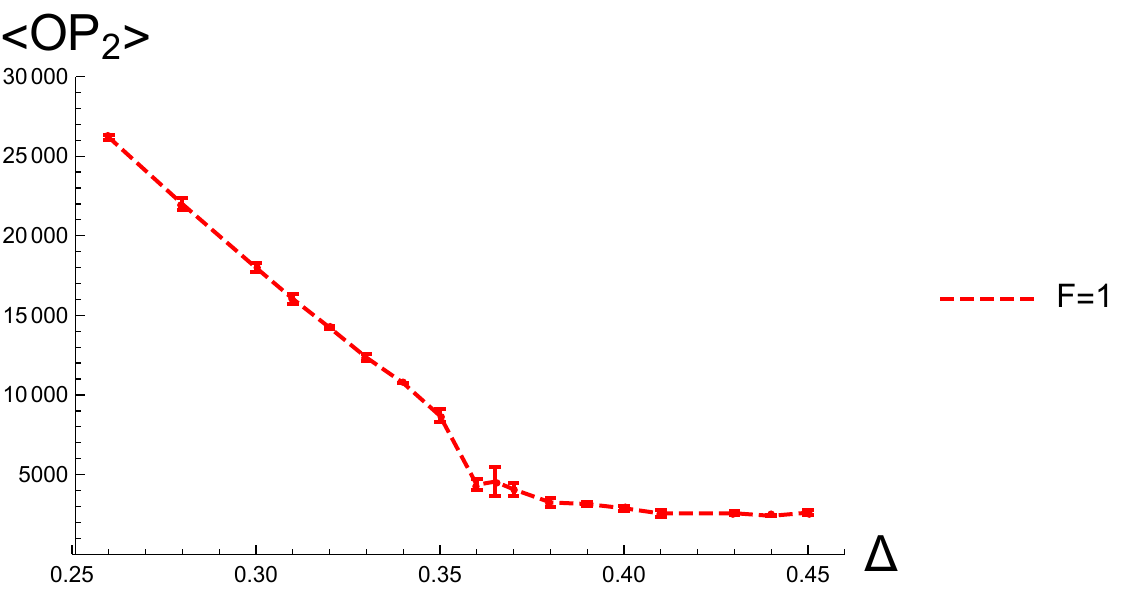}}
  \scalebox{.6}{\includegraphics{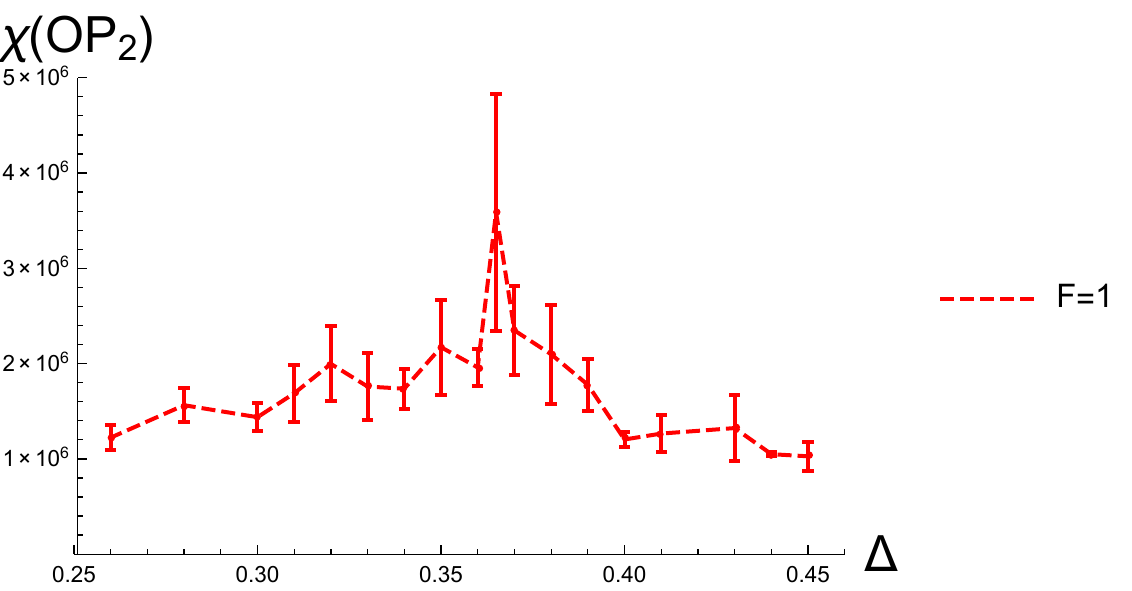}}
  \scalebox{.6}{\includegraphics{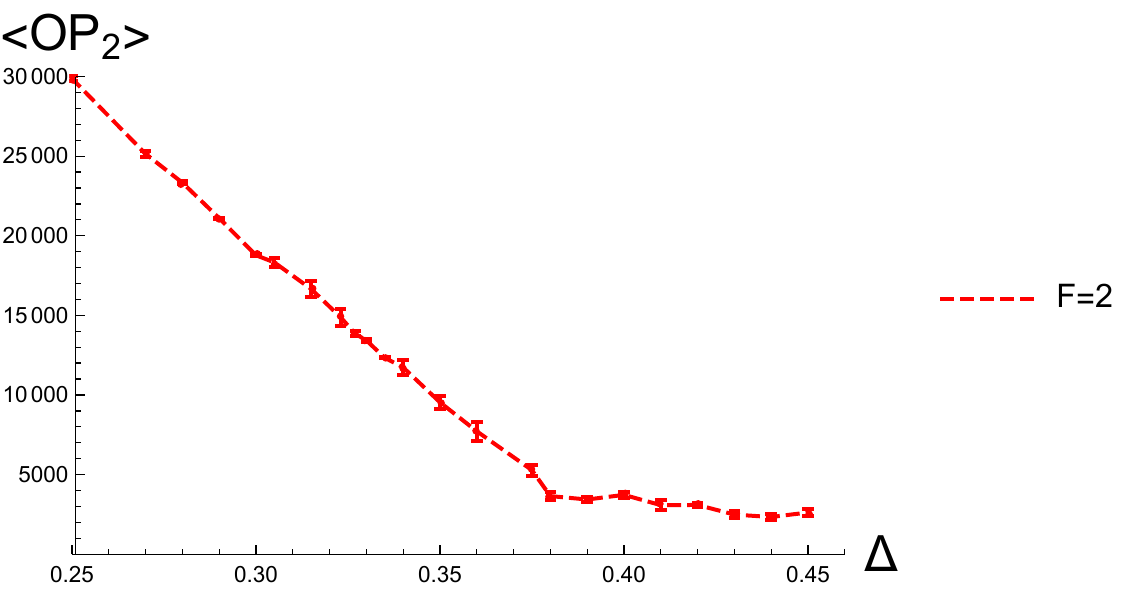}}
   \scalebox{.6}{\includegraphics{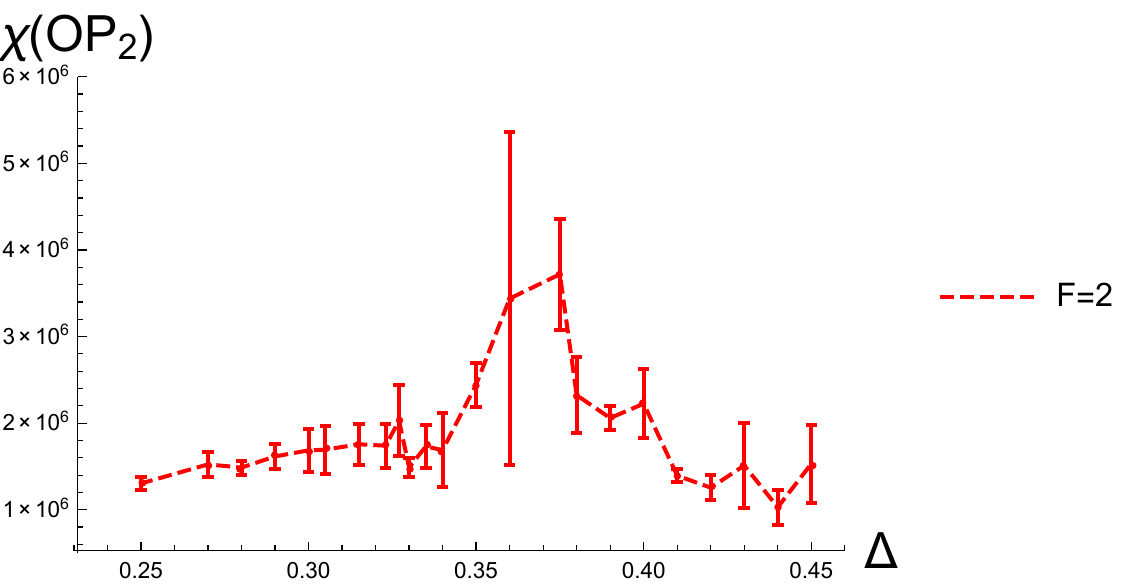}}
  \scalebox{.6}{\includegraphics{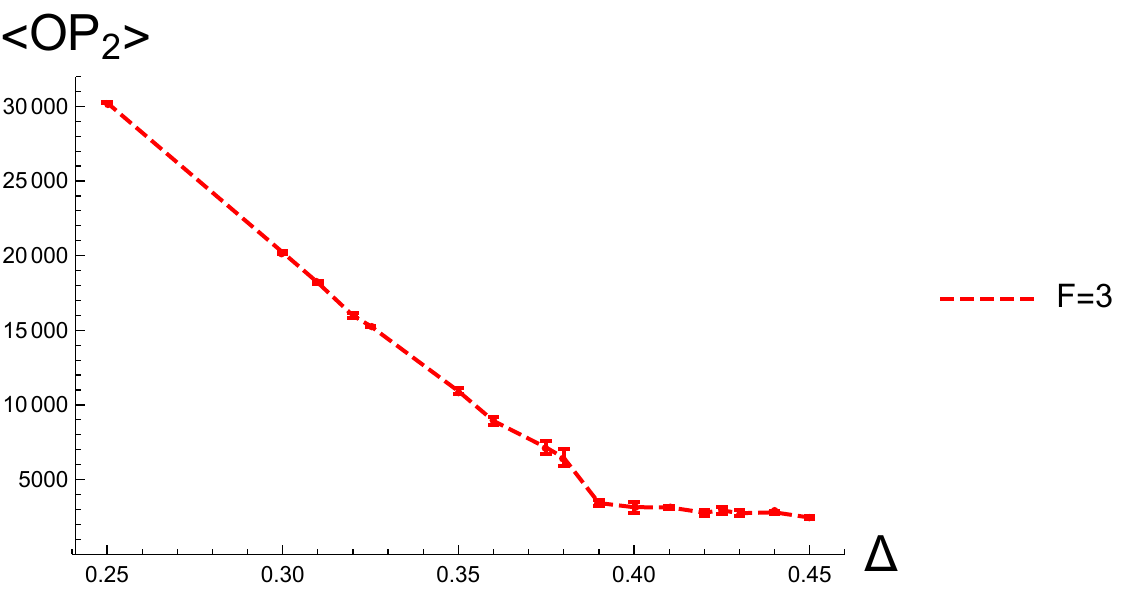}}
  \scalebox{.6}{\includegraphics{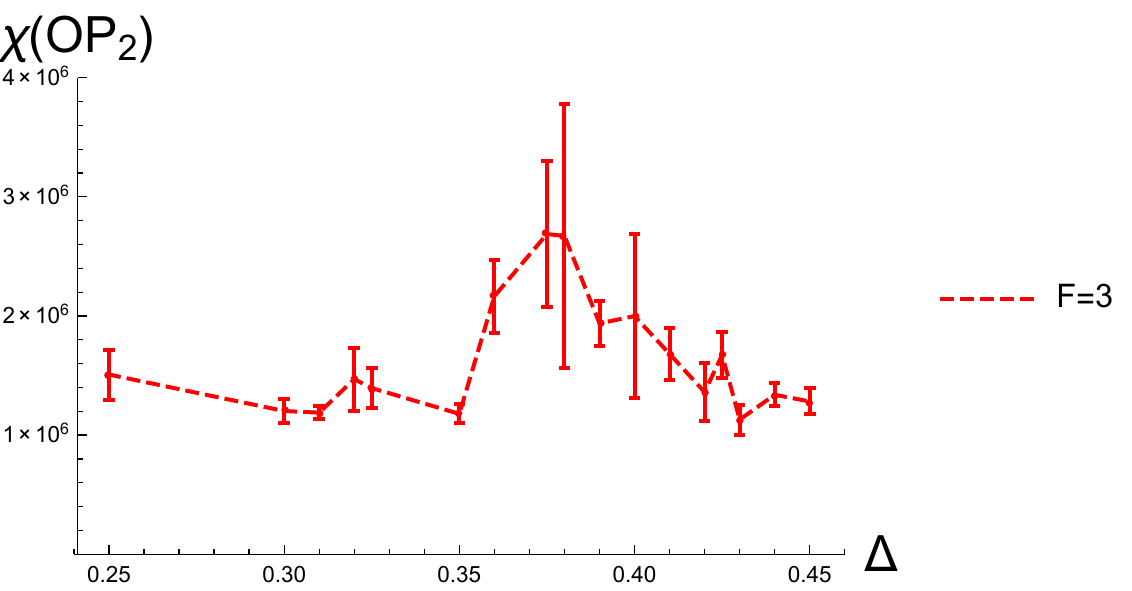}}
   \caption{\small The mean value of the order parameter $\langle \text{OP}_2 \rangle$ (left) and its susceptibility $\chi_{\text{OP}_2}$ (right) as a function of $\Delta$ for CDT including 1, 2 and 3 massless scalar fields with a lattice volume $N_{4,1}=160k$.}
\label{FigFields1}
\end{figure}

\end{section}

\begin{section}{Discussion and conclusions}\label{Discussion}

The approach of causal dynamical triangulations (CDT) has produced a number of important results, however some key questions still remain. Principle among these open problems is whether CDT has a continuum limit. An important step towards answering this question will be to determine whether there exists a second order phase transition that is accessible from within the physically interesting phase $C_{dS}$, at which point the correlation length becomes infinite so that one can keep observable quantities fixed in physical units while the lattice spacing is taken to zero. In this work we have presented strong evidence that the transition between phases $C_{dS}$ and $C_{b}$ is greater than first order, therefore presenting a strong candidate for the long sought after second order transition.

Using an order parameter that exploits the geometric differences in phases $C_{dS}$ and $C_{b}$ we are able to approximately locate the position of the (pseudo-)critical phase transition for a number of different lattice volumes $N_{4,1}$. By measuring how the position of the phase transition depends on the lattice volume $N_{4,1}$ we can extract a value for the critical exponent $\gamma$, which indicates the order of the phase transition. A first order transition is characterised by a critical exponent $\gamma=1$, whereas for a higher order transition one would expect $\gamma \neq 1$. Using 8 different lattice volumes we determine the critical exponent of the $C_{dS}-C_{b}$ transition to be $\gamma=2.71 \pm 0.34$. This result marks a significant improvement on the preliminary results found in Ref.~\cite{Ambjorn:2015qja,   Coumbe:2015oaa} and establishes that the transition is greater than first order with a $99\%$ confidence interval. This result strongly supports the conjecture that the $C_{dS}-C_b$ phase transition is a higher order transition.  
 
Motivated by Hartle and Hawking's suggestion that a sufficient number of matter fields may be a necessary condition to produce the correct classical behaviour of the universe \cite{Hartle:2008ng} we have also investigated the effect of adding $N$ massless scalar fields to the bare lattice action of CDT. Specifically, we have studied the impact of scalar fields on average spatial volume profiles and spatial volume fluctuations in the de Sitter phase $C_{dS}$ as well as the bifurcation phase $C_b$. We observe that the addition of massless scalar fields causes both the volume and fluctuation profiles to narrow in the time direction, with the same qualitative behaviour observed in phases $C_{dS}$ and $C_b$.

Using the same order parameter studied in the case of pure gravity (zero massless scalar fields) we have also analysed whether the position of the $C_{dS}-C_b $ transition line depends on the number of massless scalar fields $N$. We find that the position of the $C_{dS}-C_{b}$ transition appears to be largely independent of the number of massless scalar fields $N$, at least for $N=1,2$ or $3$. This result may be interpreted as suggesting that the bifurcation phase is probably not simply an artifact of the naive pure gravity formulation of CDT.

\end{section}


\section*{Acknowledgements}

JGS and JJ wish to acknowledge the support of the grant DEC-2012/06/A/ST2/00389 from the National Science Centre Poland. JA and DNC wish to acknowledge support from the ERC-Advance grant 291092, ``Exploring the Quantum Universe'' (EQU). AG acknowledges support by the National Science Centre, Poland
under grant no. 2015/17/D/ST2/03479.



\bibliographystyle{unsrt}
\bibliography{Master}

\end{document}